\documentclass[11pt]{article}
\usepackage{amssymb}
\usepackage{amsfonts}
\usepackage{amsthm}
\usepackage{graphicx}
\usepackage{amsmath}
\usepackage{lineno}
\usepackage[authoryear]{natbib}

\newcommand{\trho}{\rho_{\text{center}}}
\newcommand{\crit}{\rho_{\text{critical}}}

\title{Designing a Resilient Allee-Ornstein-Uhlenbeck model}

\date{}

\author{Luis F. Gordillo\\ 
{\small Department of Mathematics and Statistics, Utah State University, Logan, UT}\\
Priscilla E. Greenwood\\ 
{\small Department of Mathematics, University of British Columbia, Vancouver, BC}}

\begin{document}

\maketitle

\begin{abstract}
In stochastic population dynamics, stochastic wandering can produce transition to an absorbing state. In particular, under Allee effects, low densities amplify the possibility of population collapse. We investigate this in an Allee-Ornstein-Uhlenbeck (Allee-OU) model, that couples a bistable Allee growth equation, with demographic noise, and environmental fluctuations modeled as an Ornstein-Uhlenbeck process. This process replaces the bifurcation parameter of the deterministic Allee effect equation. In the model, small noise may induce escape from the safe basin around the positive equilibrium toward extinction.
We construct a stochastic control, altering the process to have a stationary distribution. We enable tractable control design, approximating the process by one with a stationary distribution. Two controlled models are developed, one acting directly on population size and another also modulating the environment. A threshold-based implementation minimizes the frequency of interventions while maximizing safe time.
Simulations demonstrate that the control stabilizes fluctuations around the equilibrium.
\end{abstract}

\section{Introduction}
In population ecology, C.S. Holling introduced the concept of resilience as referring to a population’s capacity to withstand disturbances, such as environmental changes or human impact, while retaining its main functions, structure and dynamic interactions among its parts (\cite{holling1973resilience}). 
In the face of escalating global environmental changes, resilience has become a cornerstone of conservation biology, encouraging strategies which safeguard vulnerable populations from tipping points that could lead to collapse (\cite{Walker2004}).
In this paper we interpret resilience as the property of a stochastic dynamical system that its paths return near a quasi-stable state when the risk of imminent extinction has crossed a threshold. 
In the context of a stochastic population process, a way to quantify resilience is via the probability that the population size remains in a desired region with low risk of extinction, for a relatively long time.
Here, we aim to maximize such a probability by using a control strategy.

A feature threatening ecological resilience is the Allee effect, where per capita fitness declines with decreasing population abundances due to mechanisms such as mate-finding failures, cooperative foraging, or inbreeding depression (\cite{CourchampBerecGascoigne2008}). In the presence of strong Allee effects, characterized by positive per capita growth rates only above a critical threshold, bistable dynamics emerge with a positive stable equilibrium, while extinction serves as an absorbing state. 
In conservation contexts, Allee effects complicate recovery efforts: reintroduced populations often fail if initial abundances fall short of the Allee threshold, as seen in efforts to bolster African wild dog (\textit{Lycaon pictus}) packs, where social cooperation thresholds amplify stochastic losses (\cite{Courchamp2001}). Moreover, in a changing climate, Allee effects may interact with habitat loss and pollution, pushing species toward critical slowing down, a precursor to abrupt shifts, emphasizing the need for proactive interventions to enhance resilience (\cite{Drake2010}). 

The situations mentioned above suggest that we study the classical equation for the strong Allee effect where the bifurcation parameter is modeled as an Ornstein-Uhlenbeck process. Here we interpret ecological resilience in the context of stochastic dynamical systems as the propensity of population paths to recover to a quasi-stable state once extinction risk surpasses a predefined threshold. Resilience can be measured quantitatively by the probability that population size persists within a ``safe" region for extended periods, minimizing the mean time to extinction (\cite{Grasman2005Resilience}). Our objective is to enhance this probability via a control strategy that embodies conservation interventions, e.g. supplementary feeding, translocation, or habitat augmentation, that nudge trajectories away from danger zones. By pulling paths back toward the safe equilibrium, controls emulate ecological management tactics that may extend population viability amid uncertainty.

Drawing on stochastic control theory (\cite{MilshteinRyashko1995, RyashkoBashkirtseva2008}), we derive feedback regulators that stabilize our system by relatively rare interventions, aligning with cost-effective conservation principles. The Wentzell-Friedlin theory is used to approximate the quasi-equilibrium process with a process having a true equilibrium, as other authors do (\cite{Bashkirtseva2011StochasticAllee,RyashkoBashkirtseva2015,Ryashko2018}), enabling design and computation of control even when the original system has an absorbing state. This tool has been used successfully in other applications (\cite{Bashkirtseva2015StochasticHH,BashkirtsevaRyashko2025NoiseTumorImmune}) and, in particular, becomes useful in our study which includes environmental control and where full stationarity is unattainable. 

The paper is structured as follows. Section 2 reviews the Allee-Ornstein-Uhlenbeck model, introducing demographic noise into the classic Allee growth equation and environmental fluctuations via an Ornstein-Uhlenbeck driven bifurcation parameter, with simulations that illustrate extinction risks. Section 3 derives a stationary law approximation  and confidence ellipses to quantify safe-region persistence. In Section 4 we construct two variants of control: one acting solely on population size (mimicking direct demographic intervention) and another also modulating the parameter process (reflecting environmental engineering). Section 5 proposes an adaptive implementation strategy, activating controls threshold-wise to minimize effort while maximizing safe time. We conclude in Section 6 with ecological implications, including ties to conservation and contrasts with early-warning paradigms.

\section{An Allee-Ornstein-Uhlenbeck model}
The Allee effect is characterized by a positive correlation between population density and the per capita population growth rate (\cite{CourchampBerecGascoigne2008}), for which several alternative mathematical models have been proposed. In particular for this paper, we adopt the commonly used scaled differential equation for the population density $x=x(t)$, 
\[\frac{dx}{dt}=x^2(1-x)-\rho x, \qquad\rho>0.\] 
This model for the Allee effect has a saddle-node bifurcation at the critical parameter value $\crit=1/4$; if $\rho>1/4$ the origin is the only stable equilibrium but if $\rho<1/4$ two additional, non-trivial equilibria appear, one unstable and other stable. See the red parabola in Figure \ref{fig: mean field}b.

We want to consider a stochastic model that has this deterministic Allee effect model as a main structure. We consider two types of stochasticity that are biologically relevant, demographic and environmental noise.
Demographic noise, which refers to random fluctuations in population size due to  births and deaths, is in general modeled by adding a multiplicative noise term to the population growth equation, which in our case is the Allee effect model, 
\begin{equation}
    dx= (x^2(1-x)-\rho x)\,dt + \sqrt{\psi x}\,dW,
    \label{eq: Allee effect and dem noise}
\end{equation}
where $\psi>0$ and $dW$ is a one dimensional Brownian motion (\cite{Dennis2016}).

In our approach, we incorporate environmental noise by modeling the bifurcation parameter corresponding to the equation of the Allee effect as an Ornstein-Uhlenbeck (OU) stochastic process (\cite{GordilloGreenwood2024}). This differs from the common practice of letting the parameters fluctuate independently at each time point as a centered Gaussian random variable with variance that is fixed or depends quadratically on the population size, e.g. (\cite{Dennis2016}).

The drift term of an OU process pulls the parameter back toward a long-term mean, and the diffusion term adds stochasticity. This makes the OU process suitable to represent environmental factors that do not jump erratically but evolve smoothly over time. Moreover, the tendency of a stochastic parameter to revert to a stable mean is ideal for modeling cases where significant deviations are temporary and the environment tends to stabilize around a typical state.
Thus $\rho=\rho(t)$ is now a solution of the stochastic differential equation 
\begin{equation}
    d\rho=-\lambda(\rho-\trho)dt+\phi\, dW,
    \label{eq: OU process equation}
\end{equation}
with $\lambda,\phi>0$ fixed. One characteristic of the process $\rho$ is that its paths fluctuate about $\trho$ with a tendency of moving towards $\trho$ that is regulated by $\lambda$. In practice, this might describe how environmental factors exhibit temporary deviations from its mean but return to it over time. In contrast to Brownian motion, the OU process avoids unbounded growth or decline, making it a reasonable model for external fluctuations.

Our Allee-OU model is obtained by coupling the Allee effect equation with demographic noise (\ref{eq: Allee effect and dem noise}) with the OU equation (\ref{eq: OU process equation}),   
\begin{equation}\label{eq: main model}
d\mathbf{X}_t =F(\mathbf{X}_t)dt+\epsilon\,\sigma(\mathbf{X}_t)
    d\mathbf{W}_t,\qquad \epsilon > 0,
\end{equation}
where
\[
\mathbf{X}_t=\begin{bmatrix}
    x(t)\\ \rho(t)
\end{bmatrix}
=\begin{bmatrix}
x\\ \rho\end{bmatrix},\quad
F(\mathbf{X}_t)=
\begin{bmatrix}
    x^2(1-x)-\rho x\\
    -\lambda (\rho-\trho)
\end{bmatrix},\quad
\sigma(\mathbf{X}_t)=\begin{bmatrix}
    \sqrt{\psi x} && 0\\ 0 && \phi
\end{bmatrix}
\]
and $d\mathbf{W}_t$ is a two dimensional Brownian motion. 

\subsection{Behavior of the field $F(\mathbf{X})$}
In what follows, let $\mathbf{X}=[x \quad \rho]'$ denote a column vector of coordinates in the phase-plane.
We begin by describing the behavior of equation (\ref{eq: main model}) in the deterministic case, i.e. $\epsilon = 0$, with $\trho < \crit$. Our focus is on the regime where $\trho$ is sufficiently close to $\crit$ that even small positive values of $\epsilon$ can produce phase transitions in the system's dynamics.

The differential equation $d\mathbf{X}/dt=F(\mathbf{X})$ has three equilibria. The first, $[0 \quad\trho]'$, which is stable, and two more, denoted by $\mathbf{X}^*_u$ and $\mathbf{X}^*_s$, which are unstable and stable, respectively, see Figure \ref{fig: mean field}(a). Figure \ref{fig: mean field}(b) shows the isoclines, determined by $F(\mathbf{X})=0$, as the parabola in red and the vertical continuous gray line. The separatrix of the system, which determines the basins of attraction of the stable equilibria, is the blue curve. Notice the region to the right of the vertical line at $\crit=0.25$ that is contained in the basin of attraction of $\mathbf{X}^*_s$. 

\subsection{Simulations of model (\ref{eq: main model})}
\begin{figure}
    \centering
    \includegraphics[scale = 0.5]{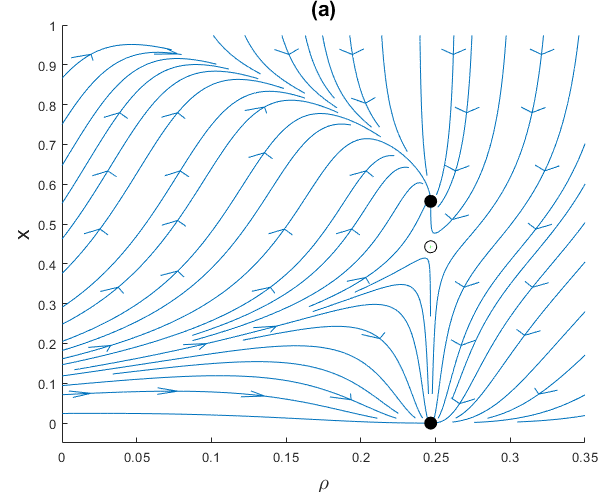}
\includegraphics[scale = 0.5]{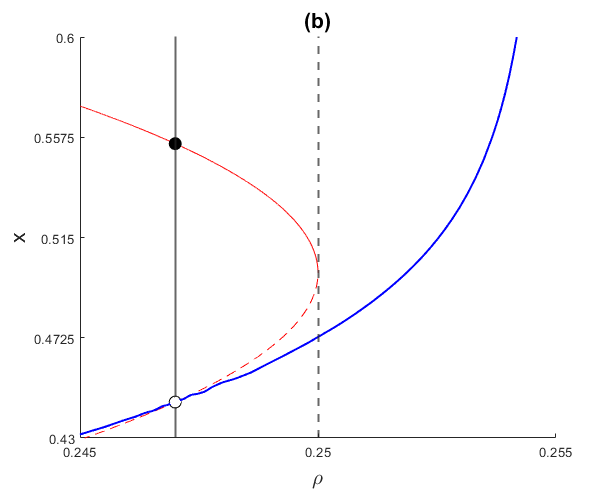}
    \caption{(a) Phase portrait for the equation $d\mathbf{X}/dt=F(\mathbf{X})$, with $\lambda=0.097$ and $\trho=0.247$. Equilibria are marked with a black (stable) and open (unstable) circles. (b) Positive equilibria and isoclines of the components of $d\mathbf{X}/dt=F(\mathbf{X})$ (red parabola and grey continuous vertical line) for $\lambda=0.01$ and $\trho=0.247$. The basin of attraction of the stable equilibrium is determined by the separatrix emerging from the unstable equilibrium (solid blue).}
    \label{fig: mean field}
\end{figure}

\begin{figure}
    \centering
    \includegraphics[scale = 0.55]{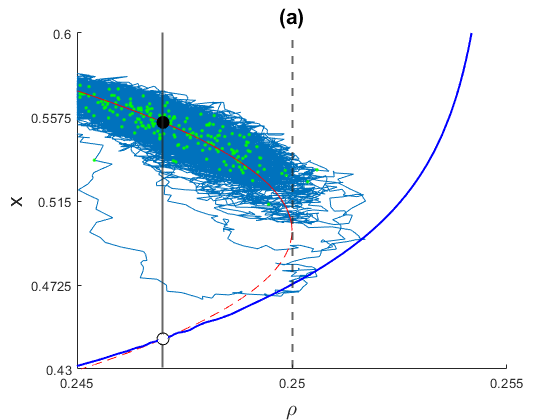}
\includegraphics[scale = 0.55]{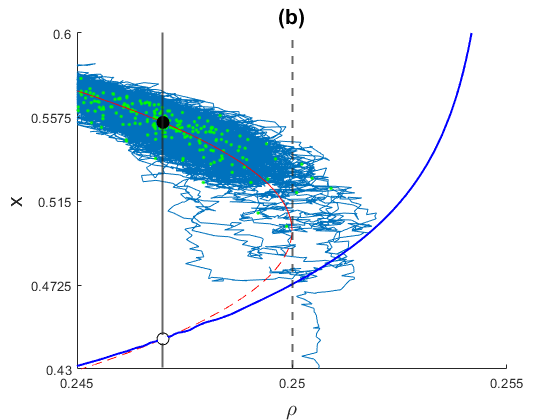}
    \caption{Sample paths of the process $\mathbf{X}_t$ starting at the stable equilibrium $\mathbf{X}_s^*$, with parameters $\lambda=0.01$,  $\trho=0.247$, $\phi=0.04$, $\psi=0.1$ and $\epsilon=0.005$. Green dots along the paths are time markers at each 150 time steps, indicating where the path lingers most of the time.}
    \label{fig: path simulations}
\end{figure}

\begin{figure}
    \centering
\includegraphics[scale = 0.55]{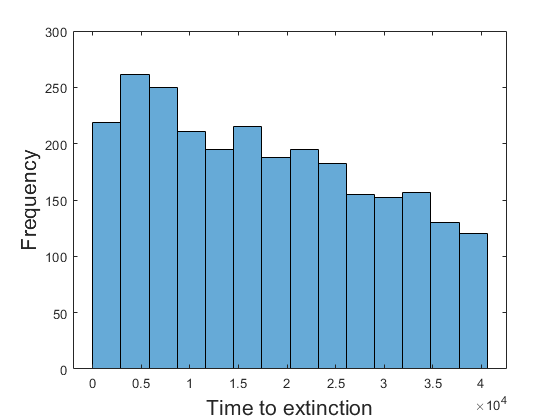}
    \caption{Histogram of times to extinction given  a fixed computational duration of $4\times 10^{4}$ time steps. Extinction time is approximated by the first crossing of the level $x=0.4$. As illustrated in Figure \ref{fig: path simulations}, sample paths may cross the separatrix and subsequently be attracted to the equilibrium $[0\quad \trho]'$, corresponding to population extinction. The parameters used are the same as in Figure \ref{fig: path simulations}. }
    \label{fig: time to extinction}
\end{figure}

Two sample paths of the process $\mathbf{X}_t$ are shown in Figure \ref{fig: path simulations}. The maximum number of time steps for the simulations is $4\times10^4$, with parameters as in Figure \ref{fig: mean field}(b), $\lambda=0.01$,  $\trho=0.247$, $\phi=0.04$, $\psi=0.1$ and $\epsilon=0.005$. 
In each panel, the path begins at $\mathbf{X}_s^*$. In (a), the path briefly crosses the separatrix but then returns to fluctuate near $\mathbf{X}_s^*$. In (b), after crossing the separatrix, the path is drawn toward the equilibrium on the horizontal axis, corresponding to the extinction of the population. The Figure indicates that the paths spend most of the time in a neighborhood of the stable equilibrium and suggests one might design a control to prevent extinction. The histogram in Figure \ref{fig: time to extinction} approximates the conditional probability distribution of the time to extinction given a fixed computational time, using the first crossing of the level $x=0.4$ as a proxy. Among 5000 simulated paths, 2630 ended in extinction. 

\section{Approximation by a model with stationary law}
In many papers, the study of a system like (\ref{eq: main model}) is done via the corresponding Fokker-Planck equation. When this equation is set equal to zero its solution is the stationary law. This is done with the hope that the answer to the question asked will not change very much, if it makes sense at all, and because computations are usually easier in the context of the stationary process. Here we are interested in computing a control which, applied to our model (\ref{eq: main model}), will cause it to be resilient against the variable environment, represented in our model by regarding the parameter $\rho$ as an OU process. We follow a line of development of \cite{RyashkoBashkirtseva2008,RyashkoBashkirtseva2015} using the stationary approximation and accompanying confidence ellipses to compute controls, which we then introduce into our system (\ref{eq: main model}) to create resilience.

For small $\epsilon>0$, the stationary distribution $p=p(\mathbf{X},\epsilon)$ concentrated around $\mathbf{X}_s^* $ satisfies 
\begin{equation}
    -\nabla \cdot F(\mathbf{X})p+\frac{\epsilon^2}{2} \left[\frac{\partial^2}{\partial x^2}S_{11}(\mathbf{X})p+\frac{\partial^2}{\partial \rho^2} S_{22}(\mathbf{X})p\right] = 0,
    \label{eq: Fokker-Planck}
\end{equation}
where $S_{11}(\mathbf{X})$ and $S_{22}(\mathbf{X})$ are the elements of the diagonal in $S(\mathbf{X})=\sigma(\mathbf{X})\sigma(\mathbf{X})'$, the apostrophe denoting the transpose. The law of the process satisfying (\ref{eq: main model}) is quasi-stationary, i.e. the probability slowly leaks across the separatrix towards the horizontal axis. 
For small $\epsilon$ we can use the approximation of order $O(1/\epsilon^2)$
\begin{equation}
    p(\mathbf{X},\epsilon) \approx M\exp\left[-\frac{Q(\mathbf{X})}{\epsilon^2} \right],
    \label{eq: approximation 1}
\end{equation}
where $M$ is a normalization constant (\cite{e2019applied, FreidlinWentzell1998})
and $Q(\mathbf{X})$ can be approximated by 
\begin{equation}
    Q(\mathbf{X})\approx \frac{1}{2}(\mathbf{X}-\mathbf{X}^*_s)'W^{-1}(\mathbf{X}-\mathbf{X}^*_s),
    \label{eq: approximation 2}
\end{equation}
where $W^{-1}$ is a positive definite matrix and with an error of order $|\mathbf{X}-\mathbf{X}_s^*|^3$ (\cite{MilshteinRyashko1995}). As shown there, the matrix $W$ satisfies
\begin{equation}\label{eq: Lyapunov equation}
    JW+WJ'+S=0,
\end{equation}
where for us $J$ is the Jacobian of the function $F$ in (\ref{eq: main model}), 
\begin{equation}
    J=
    \begin{pmatrix}
2x-3x^2-\rho & -x\\
0 & -\lambda
\end{pmatrix},
\end{equation}
evaluated at
\[\mathbf{X}^*_s=\left(\frac{1}{2}\left(1+\sqrt{1-4\trho}\right),\,\trho\right)'\]
and $S=S(\mathbf{X}^*_s)=\sigma(\mathbf{X}^*_s)\sigma(\mathbf{X}^*_s)'$.
We combine (\ref{eq: approximation 1}) and (\ref{eq: approximation 2}) to approximate $p(\mathbf{X},\epsilon)$ as a Gaussian with covariance $\epsilon^2 W$, 
\begin{equation}
    p(\mathbf{X},\epsilon)\approx M\exp\left[-\frac{(\mathbf{X}-\mathbf{X}^*_s)'W^{-1}(\mathbf{X}-\mathbf{X}^*_s)}{2\epsilon^2} \right],
\end{equation}  
when $\epsilon$ is small and during time intervals where $\mathbf{X}$ is close to $\mathbf{X}_s^*$.

\begin{figure}
    \centering
    \includegraphics[scale = 0.55]{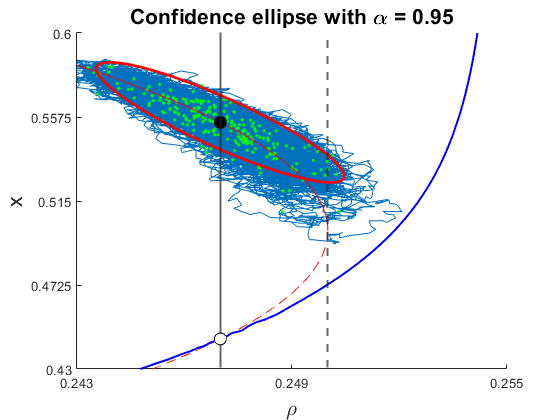}
    \caption{Confidence ellipse for the approximate model with stationary law and level of confidence $\alpha=0.95$ plotted together with a sample path of the process $\mathbf{X}_t$, solution of equation (\ref{eq: main model}), with $\mathbf{X}_0=\mathbf{X}_s^*$ and time horizon $T$. Our impression from the plot is that the approximating stationary law is a reasonable proxi for the law of the process $\mathbf{X}_t$ within the chosen $T$. Parameters for the process are the same as in Figure 2.}
    \label{fig: confidence ellipse}
\end{figure}

With the parameters chosen for our simulations in Section 2.3, the matrix $W$ is computed by solving equation (\ref{eq: Lyapunov equation}),
\[ W = \begin{bmatrix}
    6.1811 & -0.6271\\
    -0.6271 & 0.08
\end{bmatrix}.\]
We use the matrix $W$ to construct confidence ellipses centered at $\mathbf{X}_s^*$. These ellipses delineate the regions in which the values of the approximate process with stationary law are concentrated. For a fixed $\epsilon>0$, a confidence ellipse is defined by the quadratic
\begin{equation}
    (\mathbf{X}-\mathbf{X}^*_s)'W^{-1}(\mathbf{X}-\mathbf{X}^*_s) = 2\epsilon^2 q,\quad q>0,
\end{equation}
where $q$ is a chi-squared value, $q=-\ln(1-\alpha)$, $\alpha$ being the level of desired confidence (e.g. 95\%) for the values under the stationary process. See the Appendix in (\cite{greenwood1996guide}) or (\cite{johnson2007applied}) Chapter 4. In other words, $\alpha$ is the probability that a value of the process under the law $p(\mathbf{X},\epsilon)$ will be inside the ellipse, Figure \ref{fig: confidence ellipse}.
The ellipse's semi-axes are scaled by the square roots of $W$'s eigenvalues, with orientations given by its eigenvectors, centered at $\mathbf{X}_s^*$.
As shown in Figure \ref{fig: confidence ellipse}, paths of the process $\mathbf{X}_t$, solution of the system (\ref{eq: main model}), can make excursions outside the ellipse. During some of those excursions, a path might cross the separatrix and be subsequently drawn toward $[0\quad\trho]'$.  

\section{Control of the stochastic system}
When $\epsilon$ is small, sample paths of system (\ref{eq: main model}) fluctuate for some time near the nontrivial equilibrium $\mathbf{X}_s^*$. However, they eventually exit its basin of attraction, crossing the separatrix, as seen in Figure \ref{fig: path simulations}. 
The crossing may be followed by either a return to the basin of attraction of $\mathbf{X}_s^*$, as in Figure \ref{fig: path simulations}(a), or by being pulled to extinction by the stable equilibrium $[0\quad \trho]'$, Figure \ref{fig: path simulations}(b). 
How can the system (\ref{eq: main model}) be stabilized, that is, modified in real time to sustain fluctuations near $\mathbf{X}_s^*$ and prevent extinction? and can we design a strategy such that the control is used only occasionally, only when needed? As in (\cite{RyashkoBashkirtseva2015,Ryashko2018}), our modifications rely on observations of the current time values of $x$ and $\rho$.
Let $K$ be a $2\times 2$ real matrix and denote explicitly the components of the stable equilibrium as $\mathbf{X}_s^*=[x^* \quad\rho^*]'$. Then introduce a control function, $U$, a feedback regulator, defined as 
\begin{equation}\label{eq: control U}
    U(\mathbf{X})=\begin{bmatrix}
        u_1(\mathbf{X}) \\
        u_2(\mathbf{X})
    \end{bmatrix} = K(\mathbf{X}-\mathbf{X}_s^*)=
    \begin{bmatrix}
        k_{11} && k_{12}\\
        k_{21} && k_{22}
    \end{bmatrix}
    \begin{bmatrix}
        x-x^* \\
        \rho-\rho^*
    \end{bmatrix},
\end{equation}
into the main equation (\ref{eq: main model}),
\begin{equation}
    d\begin{bmatrix}
        x \\ \rho
    \end{bmatrix}
    =
    \underbrace{\left(
    \begin{bmatrix}
    x^2(1-x)-\rho x\\
    -\lambda (\rho-\trho)
    \end{bmatrix} +
    \begin{bmatrix}
        k_{11} && k_{12}\\
        k_{21} && k_{22}
    \end{bmatrix}
    \begin{bmatrix}
        x-x^* \\
        \rho-\rho^*
    \end{bmatrix}
    \right)}_{G(\mathbf{X},U):=F(\mathbf{X})+U(\mathbf{X})} dt + \epsilon
    \begin{bmatrix}
    \sqrt{\psi x} && 0\\ 0 && \phi
\end{bmatrix}
\begin{bmatrix}
    dW_1 \\
    dW_2
\end{bmatrix}.
\label{eq: controlled system}
\end{equation}
Let us denote the new drift function by $G$. Notice that, although the system has changed, the point $\mathbf{X}_s^*$ is still an equilibrium for the new drift. We want $\mathbf{X}_s^*$ to be stable, so the sample paths of the process $\mathbf{X}_t$ keep fluctuating around it. This can be achieved if the matrix $K$ belongs to the set 
\begin{equation}
\mathbb{K}=\{K\,|\,\text{ the real parts of the eigenvalues of $J+BK$ are negative} \},    
\label{eq: condition for K}
\end{equation}
where \(B=\partial G/\partial U (\mathbf{X}_s^*,0)\), is the Jacobian of the control with respect to $U$ evaluated at $(\mathbf{X}_s^*,0)$, (\cite{RyashkoBashkirtseva2008}).
We examine two scenarios: (1) the control influences both the population $x$ and the process $\rho$, and (2) the control influences only the population $x$. These correspond to distinct kinds of environment: the former represents controlled settings, such as a laboratory, while the latter reflects broader, uncontrolled scenarios, like natural environments.

\subsection{$U$ acts on $x$ and $\rho$}
Let us construct a control $U$ that acts on both components of $\mathbf{X}_t$, that is, the rank($B$) is equal to 2.
Following (\cite{RyashkoBashkirtseva2008}), the new stochastic sensitivity matrix $W$ is a solution of the equation
\begin{equation}
    (J+BK)W+W(J+BK)'+S=0,
    \label{eq: extended Liapunov equation 1}
\end{equation}
which is analogous to equation (\ref{eq: Lyapunov equation}) obtained for the system without control. The matrix $K$ can be computed as
\begin{equation}
    K = -B^{-1}\left(J+\frac{1}{2}SW^{-1}\right).
    \label{eq: relation for K and W 1}
\end{equation}
In our case, $B$ is the identity matrix $I$. As explained in (\cite{RyashkoBashkirtseva2008}), the matrix $W$ can be chosen as a multiple of the identity, $wI$, with $w$ small, which results in a matrix $K$ that belongs to $\mathbb{K}$. For our example we chose $w=0.04$, which yields the matrices
\[
W=\begin{bmatrix}
    0.04 & 0 \\
    0 & 0.04
\end{bmatrix}, \qquad
K=\begin{bmatrix}
  -0.6271 & 0.5548 \\
  0 & -0.01
\end{bmatrix}.
\]
Figure \ref{fig: dynamics with control}(a) shows a path of the controlled process, solution of (\ref{eq: controlled system}), that starts at $\mathbf{X}_s^*$, computed for the time horizon $T$. We see that the path fluctuates very close around the equilibrium. A confidence ellipse under the approximate stationary law for the controlled process appears in red.

\subsection{$U$ acts only on $x$}
\begin{figure}
    \centering
    \includegraphics[scale=0.55]{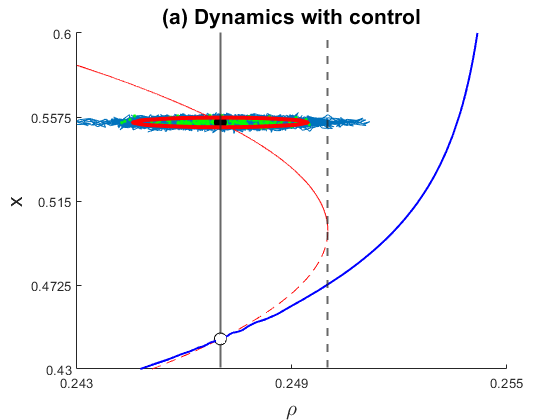}
\includegraphics[scale=0.55]{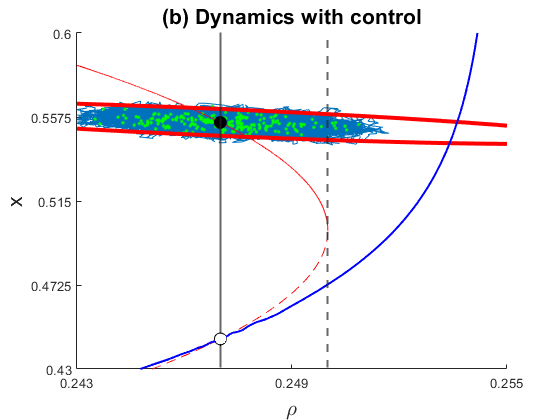}
    \caption{Sample paths of the system (\ref{eq: controlled system}) with control $U$ starting at $\mathbf{X}_s^*$ and plotted on top of Figure \ref{fig: mean field}(b) only for reference. Confidence ellipses under the approximate stationary law for the controlled process are the thick red curves. (a) $U$ acts on $x$ and $\rho$. (b) $U$ acts only on $x$.  Contrasting with Figure \ref{fig: confidence ellipse}, the controlled paths are kept tightly close to the stable equilibrium $\mathbf{X}_s^*$ in both cases.  All the parameters and the level of confidence, $\alpha$, are as in Figure \ref{fig: confidence ellipse}.}
    \label{fig: dynamics with control}
\end{figure}

Let us construct a control $U$ that acts only on $x$, i.e. $k_{21}=k_{22}=0$ in (\ref{eq: control U}), which makes rank($B$) equal to one. 
Let us write
\[ H(W)=JW+WJ'+S,\]
then, following (\cite{RyashkoBashkirtseva2008}), the matrix $W$ must satisfy
\begin{equation}
    PH(W)P=0, 
\end{equation}
where $P=I-BB^+$ and $B^+$ is the pseudoinverse of $B$. It is also shown there that the matrix $K$ can be computed using the formula
\begin{equation}
    K = -\frac{1}{2}B^+H(W)(P+I)W^{-1}.
    \label{eq: relation between K and W 2}
\end{equation}
The diagonal entries of the matrix $W$ may be assigned a common value $w>0$, provided that the condition in (\ref{eq: condition for K}) is satisfied.
The computations also show that the elements off the diagonal of $W$ coincide with those in the matrix corresponding to the uncontrolled system. In view of the constraints, for our example, a possible way of defining the matrix $W$ and the corresponding matrix $K$ are
\[
W=\begin{bmatrix}
  0.8 & -0.6271 \\
  -0.6271 & 0.8
\end{bmatrix},\qquad
K=\begin{bmatrix}
    -0.0451 & 0.4639 \\
    0 & 0
\end{bmatrix}.
\]
Figure \ref{fig: dynamics with control}(b)  shows a path of the controlled process, solution of (\ref{eq: controlled system}), that starts at $\mathbf{X}_s^*$, computed for the time horizon $T$. We see that the path fluctuates very close around the equilibrium. A confidence ellipse under the approximate stationary law for the controlled process appears in red.

\subsection{Dynamics under the controlled field $F(\mathbf{X})+U(\mathbf{X})$}
\begin{figure}
    \centering
    \includegraphics[scale=0.55]{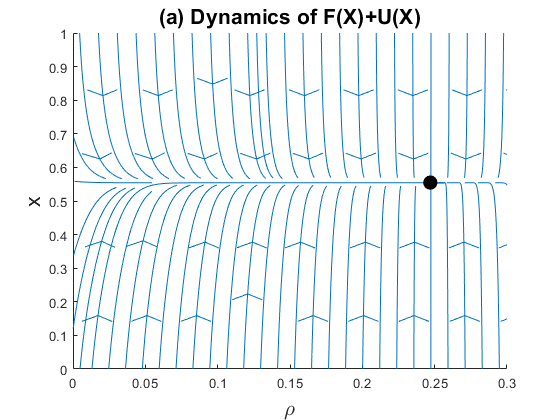}
    \includegraphics[scale=0.55]{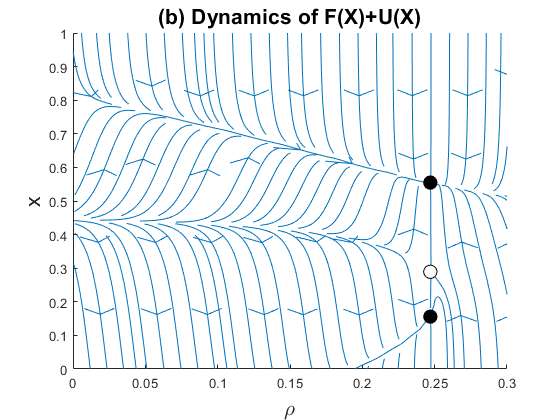}
    \caption{Phase portrait for the equation $d\mathbf{X}/dt=F(\mathbf{X})+
    U(\mathbf{X})$, with parameters $\lambda=0.01$ and $\trho=0.247$. (a) The matrix $K$ is computed as indicated in Section 4.1. (b) The matrix $K$ is computed as indicated in Section 4.2.} 
    \label{fig: Flow of F(X)+U(X)}
\end{figure}

It is straightforward to verify that the point $\mathbf{X}_s^*$, the stable equilibrium of the field without control, is also a stable equilibrium of $d\mathbf{X}/dt=F(\mathbf{X})+U(\mathbf{X})$. However, other equilibria are different. In our example, the system controlling both variables as indicated in Section 4.1 has only one real, stable equilibrium, Figure \ref{fig: Flow of F(X)+U(X)}(a), whereas when controlling only $x$ according to Section 4.2, the system has two positive stable and one unstable equilibria, Figure \ref{fig: Flow of F(X)+U(X)}(b). 
Although in both cases stochasticity will eventually cause any path that starts close to $\mathbf{X}_s^*$ to move against the deterministic flow and to hit the horizontal axis with probability one, the time for this to happen is relatively large in comparison with that of the model without the control. For instance, we repeated 5000 simulations for each type of control and in none of them did the path reach extinction during the same time frame used for generating the histogram in Figure \ref{fig: time to extinction}.

\section{A strategy for implementing the control}
\begin{figure}
    \centering
    \includegraphics[scale = 0.6]{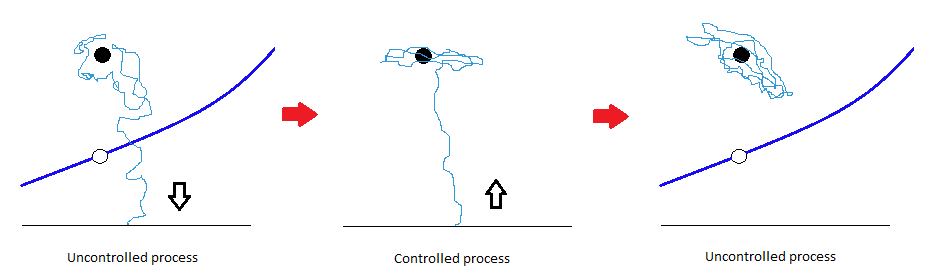}
    \caption{Strategy for implementing the control. \textbf{Left:} A path of the uncontrolled process fluctuates around $\mathbf{X}_s^*$ (black dot). It eventually crosses the separatrix (blue curve) and is pulled down until it meets a predetermined horizontal level (horizontal black line). \textbf{Center:} The controlled process activates, drawing the path upward toward the equilibrium $\mathbf{X}_s^*$, where it fluctuates for a short time. \textbf{Right:} Switch back to the uncontrolled process.}
    \label{fig: strategy for control}
\end{figure}

Our proposed strategy combines the controlled and uncontrolled processes, creating a new process with a stationary law, centered around $\mathbf{X}_s^*$. Paths of the new process will fluctuate near $\mathbf{X}_s^*$ with high probability. The idea is to replace the uncontrolled process with the controlled one whenever the risk of extinction becomes high, thereby driving the system back to a neighborhood of $\mathbf{X}_s^*$. Once the trajectory is near this equilibrium, the dynamics switches back to the uncontrolled process, creating an alternating pattern. Figure \ref{fig: strategy for control} shows a schematic of the procedure, where the control is implemented when the path of the population process crosses a fixed level. 
This procedure creates a stationary process. The next aim would be to maximize the time spent near $\mathbf{X}_s^*$  and minimize the use of the control. 

\begin{figure}
    \centering
    \includegraphics[scale = 0.55]{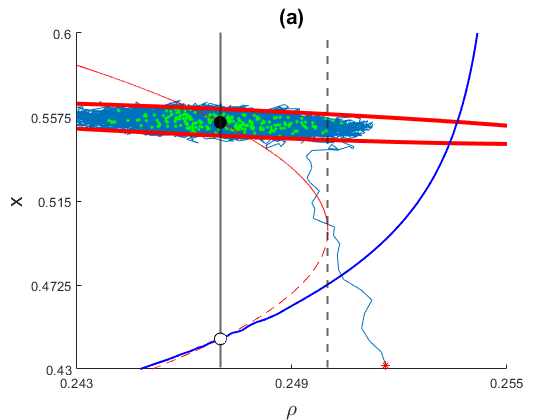}
    \includegraphics[scale = 0.55]{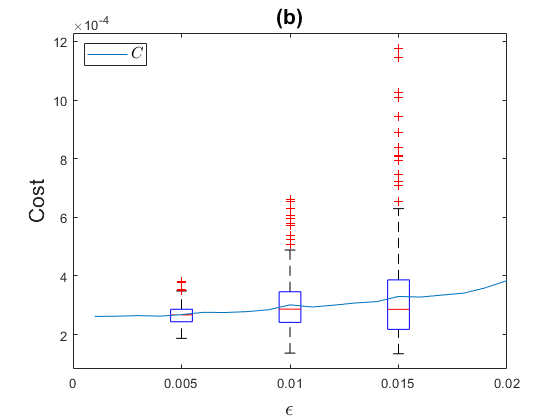}
    \includegraphics[scale = 0.55]{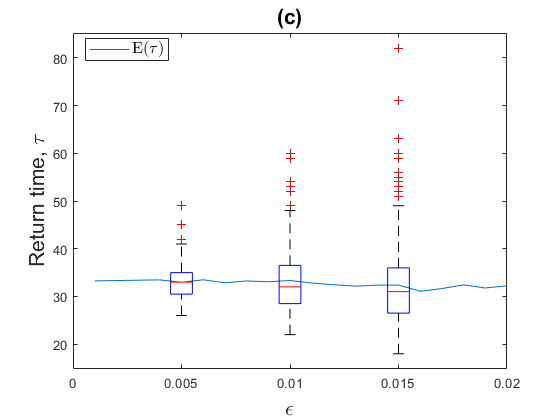}
    \caption{(a) A path of the controlled process representing the intermediate stage of the extinction-avoidance strategy illustrated in Figure \ref{fig: strategy for control}. (b) The control cost and boxplots, as function of $\epsilon$. (c) Mean and boxplots of the return time $\tau$ (first passage time) through the boundary $x=x^*-0.005$.   
    }
    \label{fig: stabilization and cost}
\end{figure}

The criteria chosen to activate the control would depend on the cost of implementing the system (\ref{eq: controlled system}). Let us consider the cost of the control as described in Section 4.2, that is, with $U$ acting only on the population process $x$. Figure \ref{fig: stabilization and cost}(a) shows a path of such a controlled process starting at an arbitrary point on the line defined by $x=0.432$, with $\rho$ between 0.243 and 0.255, which quickly stabilizes around $\mathbf{X}_s^*$. We call this point the ``initial point of return". 
The control in this case is
\[
U(\mathbf{X}_t) = \begin{bmatrix} u_1 (\mathbf{X}_t)\\ 0 \end{bmatrix}, \quad u_1(\mathbf{X}_t) = k_{11} (x - x^*) + k_{12} (\rho - \trho).   
\]
In classical stochastic control theory (e.g., \cite{FlemingRishel1975}), the cost of implementing a feedback control $U(\mathbf{X}_t)$ is quantified through an expected infinite-horizon quadratic cost functional. This functional penalizes deviations of the state variables, i.e. cost of stabilizing the system near the equilibrium, and the effort required to apply the control. We set our control cost $C$ considering only penalties of the former type. We define $C$ as  
\begin{equation}
C = \mathbb{E} \left[ \int_0^\tau r \left[ k_{11} (x - x^*) + k_{12} (\rho - \trho) \right]^2 \, dt \right],
\label{eq: control cost} 
\end{equation}
where $r>0$ is the relative penalty for the control $ u_1 $ and $\tau$ is the returning time (first passage time) from the initial point of return through the barrier defined by $x=x^*-\eta$, $\eta>0$ small. The choice of $\eta$ must ensure that, with high probability, beyond this boundary the path resumes fluctuating around $\mathbf{X}_s^*$ when the control is removed.

The cost function is based on deviations of $x(t)$ from $x^*$ and $\rho(t)$ from $\trho$. The term $k_{12} (\rho - \trho)$ is the cost for responding to environmental fluctuations. 
Figure \ref{fig: stabilization and cost}(b) shows the control cost, $C$, of moving the process from the initial point of return to the first crossing of the horizontal line $x=x^*-0.005$. For each value of $\epsilon$, the cost $C$ is estimated as the average of 300 simulations, with $r=1$, and the initial point of return, as stated before, with $x(0)$ on the line defined by $x=0.432$ and $\rho_0$ drawn uniformly in $[0.235,0.255]$.
The plot shows that the control cost $C$ barely increases with increasing $\epsilon$ but its variability increases.

Figure \ref{fig: stabilization and cost}(c) shows the mean of the return time $\tau$ (first passage time) through the boundary  $x=x^*-0.005$, which slightly decreases with increasing $\epsilon$, whereas its variability increases.

\section{Conclusions and discussion}

In this paper we propose a way to construct resilience in stochastic population models exhibiting Allee effects. 
We use a stationary distribution approximation to derive a strategy that ideally eliminates extinction risks. The novelty here is the computation of controls for the Allee-OU model.
Our simulations reveal that these controls can largely confine fluctuations of the process around the safe equilibrium $\mathbf{X}_s^*$, reducing extinction probabilities from over 50\% in uncontrolled paths to near zero in controlled regimes over a simulated time horizon. 

Although we do not address the problem of how to execute the control, our control design offers a theoretical tool for proactive conservation in Allee-afflicted species, where low densities may precipitate failures in fitness components such as mating or foraging cooperation. The control executed only on the population ($U$ acting only on $x$) suggests direct interventions such as supplementary feeding or translocations, which artificially increase the density toward a safe region. In contrast, the dual control ($U$ acting on both $x$ and $\rho$), echoes environmental engineering tactics, such as habitat restoration or targeted predator removal to revert fluctuations toward favorable states.  Our threshold-activated implementation triggers interventions only when paths cross a predefined risk boundary. Our simulations show that the time spent by the process around the equilibrium $\mathbf{X}_s^*$ between the applications of the control is large in comparison to the time spent under the control, see Figures \ref{fig: time to extinction} and \ref{fig: stabilization and cost}(c).

Our two-dimensional, non-spatial model does not include the effects of dispersal among population patches, which might induce secondary effects, such as ``emergent resilience". This refers to the possibility of dispersal acting as a stabilizing force by importing individuals from other source patches, countering local extinctions, a ``rescue effect". This can make the overall metapopulation more robust to noise than isolated patches would suggest. 

Our approach of dynamic stabilization contrasts with early-warning signals that detect imminent collapses via precursors such as increased variance or autocorrelation (\cite{Drake2010}). While early-warning signals excel in forecasting demographic regime shifts, our controls prevent such transitions by reshaping the stochastic dynamics landscape. 

Future extensions could incorporate spatial heterogeneity and multi-species interactions, e.g., predator-prey Allee cascades. We hope that our construction contributes to the transformation of theoretical resilience understanding to actionable conservation.
\\
\\
\textbf{Data availability.} MATLAB codes used to generate the figures are available at \\https://github.com/LuisFGordillo/Codes

\bibliographystyle{plainnat}
\bibliography{refs}

@article{Bashkirtseva2011StochasticAllee,
  author  = {I. Bashkirtseva and L.B. Ryashko},
  title   = {Sensitivity analysis of stochastic attractors and noise-induced transitions for population model with {A}llee effect},
  journal = {Chaos},
  year    = {2011},
  volume  = {21},
  number  = {4},
  pages   = {047514},
  doi     = {10.1063/1.3647316}
}

@article{Bashkirtseva2015StochasticHH,
  author    = {I. Bashkirtseva and A.B. Neiman and L.B. Ryashko},
  title     = {Stochastic sensitivity analysis of noise-induced suppression of firing and giant variability of spiking in a {H}odgkin–{H}uxley neuron model},
  journal   = {Physical Review E},
  volume    = {91},
  number    = {5},
  pages     = {052920},
  year      = {2015},
  doi       = {10.1103/PhysRevE.91.052920},
  url       = {https://doi.org/10.1103/PhysRevE.91.052920}
}

@article{BashkirtsevaRyashko2025NoiseTumorImmune,
  author  = {I. Bashkirtseva and L. Ryashko},
  title   = {Positive and negative role of random perturbations in the dynamics of a tumor-immune system with treatment},
  journal = {Journal of Mathematical Biology},
  year    = {2025},
  volume  = {91},
  number  = {2},
  pages   = {20},
  doi     = {10.1007/s00285-025-01694-y},
  epub    = {2025-07-26},
  pmid    = {40715848}
}

@article{Courchamp2001,
author = {Courchamp, F. and Macdonald, D.W.},
title = {Crucial importance of pack size in the African wild dog Lycaon pictus},
journal = {Animal Conservation},
volume = {4},
pages = {169--174},
year = {2001},
doi = {}
}

@book{CourchampBerecGascoigne2008,
  author    = {Courchamp, F. and Berec, L. and Gascoigne, J.},
  title     = {Allee Effects in Ecology and Conservation},
  publisher = {Oxford University Press},
  address   = {Oxford; New York},
  year      = {2008},
  series    = {Oxford Biology},
  pages     = {x, 256},
  isbn      = {9780198570301},
  doi       = {10.1093/acprof:oso/9780198570301.001.0001},
}

@article{Dennis2016,
  author       = {Dennis, B. and Assas, L. and Elaydi, S. and Kwessi, E. and Livadiotis, G.},
  title        = {Allee effects and resilience in stochastic populations},
  journal      = {Theoretical Ecology},
  volume       = {9},
  number       = {3},
  pages        = {323--335},
  year         = {2016},
  doi          = {10.1007/s12080-015-0288-2},
}

@article{Drake2010,
author = {Drake, J.M. and Griffen, B.D.},
title = {Early warning signals of extinction in deteriorating environments},
journal = {Nature},
volume = {467},
number = {7314},
pages = {456--459},
year = {2010},
doi = {10.1038/nature09389},
publisher = {Macmillan Publishers Limited}
}

@book{e2019applied,
  title        = {Applied Stochastic Analysis},
  author       = {E, W. and Li, T. and Vanden-Eijnden, E.},
  series       = {Graduate Studies in Mathematics},
  volume       = {199},
  year         = {2019},
  publisher    = {American Mathematical Society},
  address      = {Providence, RI},
  isbn         = {978-1-4704-6569-8}
}

@book{FlemingRishel1975,
  author    = {Fleming, W.H. and Rishel, R.W.},
  title     = {Deterministic and Stochastic Optimal Control},
  publisher = {Springer-Verlag},
  address   = {Berlin, Heidelberg},
  year      = {1975},
  doi       = {10.1007/978-3-642-10942-4},
  isbn      = {978-3-642-10944-8}
}

@book{FreidlinWentzell1998,
  author       = {Freidlin, M.I. and Wentzell, A.D.},
  title        = {Random Perturbations of Dynamical Systems},
  edition      = {2},
  series       = {Grundlehren der Mathematischen Wissenschaften},
  volume       = {260},
  publisher    = {Springer},
  address      = {New York, NY},
  year         = {1998},
  isbn         = {978-3-540-90858-6},
}

@article{GordilloGreenwood2024,
  author       = {Gordillo, L.F. and Greenwood, P.E.},
  title        = {A stochastic mechanism causing long or short transients near a bifurcation point},
  journal      = {Proceedings of the Royal Society A},
  volume       = {480},
  year         = {2024},
  number       = {20240329},
  doi          = {10.1098/rspa.2024.0329},
}

@inbook{Grasman2005Resilience,
  author       = {J. Grasman and O.A. van Herwaarden and T.H.J. Hagenaars},
  title        = {Resilience and persistence in the context of stochastic population models},
  booktitle    = {Current Themes in Theoretical Biology: A Dutch Perspective}, 
  publisher    = {Springer},
  year         = {2005},
  pages        = {267--280},
  isbn         = {9781402029011},
  doi          = {10.1007/1-4020-2904-7\_10}
}

@book{greenwood1996guide,
  author    = {Greenwood, P.E. and Nikulin, M.S.},
  title     = {A Guide to Chi-Squared Testing},
  year      = {1996},
  publisher = {John Wiley \& Sons},
  address   = {New York},
  series    = {Wiley Series in Probability and Statistics},
  
  isbn      = {0-471-55779-X, 978-0471557791},
  pages     = {xii + 280}
}

@article{holling1973resilience,
  title={Resilience and stability of ecological systems},
  author={Holling, C.S.},
  journal={Annual Review of Ecology and Systematics},
  volume={4},
  pages={1--23},
  year={1973},
  publisher={Annual Reviews},
  doi={10.1146/annurev.es.04.110173.000245}
}

@book{johnson2007applied,
  title     = {Applied Multivariate Statistical Analysis},
  author    = {Johnson, R.A. and Wichern, D.W.},
  year      = {2007},
  edition   = {6th},
  publisher = {Pearson Prentice Hall},
  address   = {Upper Saddle River, NJ},
  isbn      = {9780131877153}
}

@article{MilshteinRyashko1995,
  author       = {Mil’shtein, G.N. and Ryashko, L.B.},
  title        = {A first approximation of the quasipotential in problems of the stability of systems with random non-degenerate perturbations},
  journal      = {Journal of Applied Mathematics and Mechanics},
  volume       = {59},
  number       = {1},
  pages        = {47--56},
  year         = {1995},
  doi          = {10.1016/0021-8928(95)00006-B},
}

@article{Ryashko2018,
  author       = {Ryashko, L.},
  title        = {Stochastic Control in the Problem of Preventing Ecological Catastrophes},
  journal      = {IFAC‑PapersOnLine},
  volume       = {32},
  number       = {51},
  pages        = {540--544},
  year         = {2018},
  doi          = {10.1016/j.ifacol.2018.11.478},
}

@article{RyashkoBashkirtseva2008,
  author       = {Ryashko, L.B. and Bashkirtseva, I.A.},
  title        = {On control of stochastic sensitivity},
  journal      = {Automation and Remote Control},
  volume       = {69},
  number       = {7},
  pages        = {1171--1180},
  year         = {2008},
  doi          = {10.1134/S0005117908070084},
}

@article{RyashkoBashkirtseva2015,
  author       = {Ryashko, L. and Bashkirtseva, I.},
  title        = {Stochastic sensitivity analysis and control for ecological model with the {A}llee effect},
  journal      = {Mathematical Modelling of Natural Phenomena},
  volume       = {10},
  number       = {2},
  pages        = {130--140},
  year         = {2015},
  doi          = {10.1051/mmnp/201510209},
}

@article{Walker2004,
author = {Walker, B. and Holling, C.S. and Carpenter, S.R. and Kinzig, A.},
title = {Resilience, Adaptability and Transformability in Social-Ecological Systems},
journal = {Ecology and Society},
volume = {9},
number = {2},
pages = {5},
year = {2004},
url = {http://www.ecologyandsociety.org/vol9/iss2/art5}
}

\end{document}